\documentclass[a4paper,10pt]{article}
\usepackage{graphicx}
\usepackage{amssymb}
\usepackage{amsmath}
\usepackage{amsfonts}
\usepackage{amsthm}

\usepackage{nicefrac}
\usepackage{dcolumn}
\usepackage{multirow}
\usepackage{cite}
\usepackage{mathrsfs}
\usepackage{bm}
\usepackage[usenames,dvipsnames]{xcolor}
\usepackage[colorlinks=true,citecolor=Blue,linkcolor=RubineRed,urlcolor=Blue]{hyperref}
\usepackage{bbm}
\usepackage{hyperref}
\usepackage{comment}
\usepackage{float}
\usepackage{caption}
\usepackage{subcaption}

\usepackage{xcolor}
\usepackage{soul}
\setstcolor{red}

\begin{document}

\huge

\begin{center}
On the electrons really contributing to dc conductivity of warm dense matter
\end{center}

\vspace{0.5cm}

\large

\begin{center}
Nadine Wetta$^{a,}$\footnote{nadine.wetta@cea.fr} and Jean-Christophe Pain$^{a,b}$
\end{center}

\normalsize

\begin{center}
\it $^a$CEA, DAM, DIF, F-91297 Arpajon, France\\
\it $^b$Universit\'e Paris-Saclay, CEA, Laboratoire Mati\`ere sous Conditions Extr\^emes,\\
\it F-91680 Bruy\`eres-le-Ch\^atel, France
\end{center}

\vspace{0.5cm}

\begin{abstract}
Atomic properties of warm dense matter is an active field of research. Understanding transport properties of these states is essential for providing coefficients needed by magneto-radiative hydrodynamics codes for many studies, including hydrodynamic instabilities, energy balances or heating in fusion plasmas, difficult to investigate by experimental means. In this paper, we present an average-atom approach for the calculation of direct-current electric conductivity within Ziman's theory. The mean ion charge $Z^*$, commonly called ionization, is an important input of the Ziman formula, but is not clearly defined within average-atom models. Our study spans a wide range of thermodynamical conditions, \emph{i.e.}, for the densities, from a few $10^{-2}$ to about 4 times the solid's density, and, for the temperatures, typically from 0.1 eV to 700 eV, favorable to large differences in the mean ion charge $Z^*$ according to its definition. We compare and discuss different ways of defining $Z^*$ while trying to figure out which electrons really contribute to electric conduction. We compare our results with experimental data and published theoretical values, in particular from the second transport code comparison workshop, which was held in July 2023 at Lawrence Livermore National Laboratory. These comparisons lead us to propose indicators for the relevance of including different charges predicted by our average-atom model in the definition of $Z^*$.
\end{abstract}

\section{Introduction}\label{sec1}

The so-called warm dense matter (WDM) regime represents a particular state of matter at the crossroads of condensed matter, plasma, and dense liquid. WDM plays a major role in many fields of physics, such as exoplanet interiors, inertial confinement fusion, and neutron-star atmospheres. It is actually as little known as it is difficult to define. The regime is inherently challenging because its thermodynamics cannot be framed in terms of small perturbations from ideal solvable models.
 
Experimental facilities now enable one to create and diagnose well-controlled WDM conditions, offering an unprecedented window on its atomic properties and providing an ideal testing ground for theoretical models. For instance, X-ray free-electron lasers (XFEL), such as the Linac Coherent Light Source (LCLS), enable one to create and study WDM conditions. The electron densities, temperatures, and ionization states of compressed materials can now be measured by means of X-ray scattering with a good spectral resolution.

The main difficulty for the modeling of WDM stems from the fact that it belongs neither to the ``traditional'' condensed-matter physics, nor to plasma physics. Plasma physics methods do not extend well from the classical limit to the WDM regime, where quantum effects are significant for the electrons, and must therefore be taken into account. Conversely, it is definitely worth extending condensed-matter physics methods beyond their normal realms of application.

An interesting possibility to describe WDM consists in combining electronic-structure calculations based on density functional theory (DFT) and molecular dynamics (MD). In its simplest form, the Born-Oppenheimer MD provides the electronic forces on the nuclei from a DFT computation and then moves the ions solving the classical Newton equations of motion, combining in that way the classical contributions of the nuclei with the quantum-mechanical treatment of the electrons. However, although relevant in the low-temperature condensed-matter phase, at higher temperatures the DFT-MD becomes prohibitive.

An alternative consists in resorting to average-atom (AA) calculations. The quantum average-atom model \cite{Rozsnyai1972,Liberman1979} is a generalization of the original finite-temperature semi-classical Thomas–Fermi model \cite{Feynman1949}. Quantum AA model can be declined in different versions, but usually relies on a self-consistent DFT calculation of the atomic structure, including exchange-correlation (XC) effects in the local density approximation (LDA) at finite temperature. In most cases, convergence is not an issue, especially as the temperature increases, and the AA model provides the equation of state as well as transport coefficients over a wide range of thermodynamic conditions.

In this article, we present calculations of electrical resistivities for beryllium, boron, carbon, aluminum, copper and gold plasmas within the Ziman-Evans theory \cite{Ziman1961}, in the framework of our average-atom code {\sc Paradisio} \cite{Penicaud2009}. The mean ion charge $Z^*$, also called ionization, is a key quantity in Ziman's theory. Different definitions are possible for $Z^*$ within AA models, which can lead to large differences in conductivities. In our study, the densities range from $\rho_\mathrm{solid}$ to a few times $\rho_\mathrm{solid}$ ($\rho_\mathrm{solid}$ being the normal density of the solid), and temperatures from 1 eV to hundreds of eV. In the case of copper, we extend our investigations to the liquid state: the densities are then slightly lower than the solid's one and temperatures vary from melting one (1350 K) to 3500 K. The diversity of the investigated thermodynamic conditions, as well as the one of the chosen elements, from simple metals to transition ones, is favorable to large differences in the mean ion charge value $Z^*$, according to its definition. We propose and discuss different ways of defining the mean ionization, while trying to figure out which electrons are really conduction ones. We compare our results with experimental data (resistivities of liquid copper) and with published theoretical values (conductivities and mean ionizations), in particular from the second charged-particle-transport code-comparison workshop, which was held in July 2023 in Livermore.

Our AA Ziman-Evans model for the calculation of electrical conductivity is briefly recalled in Sec. \ref{sec2}. The identification of conduction electrons and the characterization of the mean ionization entering the Ziman formula are discussed in Sec. \ref{sec3}. Comparisons with mean ionizations inferred from QMD (Quantum Molecular Dynamics) simulations, through Hugoniot curves and dynamical conductivities, are presented in Sec. \ref{sec4}. Our results are compared with data (DFT-MD and AA calculations) from the July 2023 second charged-particle transport coefficient code comparison workshop in Sec.~ \ref{sec5}. Our AA model describes an atom confined in a finite spherical volume (Wigner-Seitz cell) surrounded by a uniform electron gas (UEG). This assumption gives rise to deviations from the Friedel sum rule, which are mostly of little consequence on conductivity calculations. In Sec. \ref{sec6}, we question the issue of the charge neutrality of the plasma, in relation with this assumption. We stress the need to compensate for the corresponding ``missing charge'' in some cases, such as for ``d-'' (or ``f-'')block liquid metals. Links with the persistence of long-range order in the liquid phase are highlighted. Our final proposal for defining the average ionic charge is based on the assumption that the sum rule is approximately verified, and that the missing charge remains small compared with the total charge $Z^*$. We therefore propose two indicative parameters to assess whether these two conditions are met.

\section{Ziman electrical resistivity in the framework of the average-atom model}\label{sec2}

\subsection{Ziman formula}\label{sec21}

Although this is not necessary under the thermodynamic conditions considered in this work, the following formulas will be given in the relativistic formalism, for the sake of consistency with the relativistic AA code {\sc Paradisio} \cite{Penicaud2009} that was used to provide the inputs required for resistivity calculations. All formulas will be given in atomic units (\emph{i.e.}: $e=\hbar=m_e=1$).\\
\indent The Ziman formulation for the electrical resistivity \cite{Ziman1961} describes, within the linear response theory, the acceleration of free electrons in a metal and their scattering by an ion. The resistivity then reads
\begin{equation}\label{eta}
    \eta=-\dfrac{1}{3\pi {Z^*} n_e} \int_0^\infty \dfrac{\partial f}{\partial \epsilon}(\epsilon,\mu^*) \mathcal{I}(\epsilon) \mathrm{d}\epsilon,
\end{equation}
where $n_e$ is the electron density, and $Z^*$ the mean ionic charge. Imposing plasma charge neutrality: $n_e=Z^* n_i,$ and $Z^* n_e={Z^*}^2 n_i,$ where we have introduced the ion number density $n_i.$ The Fermi-Dirac distribution and its derivative respectively read
\begin{equation}\label{fermidirac}
    f(\epsilon,\mu^*)=\dfrac{1}{e^{\beta (\epsilon-\mu^*)}+1}
\end{equation}
and
\begin{equation}
    \dfrac{\partial f}{\partial\epsilon}(\epsilon,\mu^*)=-\beta f(\epsilon,\mu^*)\left[1-f(\epsilon,\mu^*)\right], 
\end{equation}
where $\beta=1/(k_B T)$ and $\mu^*$ denotes the chemical potential associated with the free electron gas of density $n_e=Z^* n_i$, given by
\begin{equation}
    \dfrac{2}{(2\pi)^3} \int_0^\infty f(\epsilon,\mu^*)\,4\pi k^2 dk=n_e,
\end{equation}
or
\begin{equation}
    \mathscr{F}_{1/2}(\beta\mu^*)=\dfrac{\pi^2}{\sqrt{2}}\beta^{3/2}n_e,
\end{equation}
where we introduced the Fermi function of order $1/2$ 
\begin{equation}
    \mathscr{F}_{1/2}(x)=\int_0^\infty \dfrac{t^{1/2}}{1+e^{t-x}}\mathrm{d}t.
\end{equation}
The function $\mathcal{I}(\epsilon)$ is given by
\begin{equation}
    \mathcal{I}(\epsilon)=\int_0^{2k}q^3 S(q) \Sigma(q) \mathrm{d}q,
\end{equation}
where $S(q)$ denotes the static ion-ion structure factor and $\Sigma(q)$ the electron-ion scattering cross-section. The vector $\vec{q}=\vec{k}^\prime-\vec{k}$ is the momentum transferred in the elastic scattering event (\emph{i.e.} such as $|\vec{k}^\prime|=|\vec{k}|$) of a conduction electron from an initial state $\vec{k}$ to a final $\vec{k}^\prime$ one. Introducing the scattering angle $\theta\equiv (\vec{k},\vec{k}^\prime$) and its cosine $\chi=\cos\theta$, one has $q^2=2k^2 (1-\chi)$ and
\begin{equation}
    \mathcal{I}(\epsilon)=2k^4 \int_{-1}^1 S\left[k\sqrt{2(1-\chi)}\right]|a(k,\chi)|^2 (1-\chi) \mathrm{d}\chi.
\end{equation}
Energy $\epsilon$ and momentum $k$ are related (within the relativistic formalism, $c$ being the speed of light, and using atomic units) by
\begin{equation}
    k=\sqrt{2\epsilon \left(1+\dfrac{\epsilon}{2c^2}\right)}.
\end{equation}
The $t-$matrix formalism of Evans \cite{Evans1973} provides the modulus of the electron-ion scattering amplitude $|a(k,\chi)|$, whose square is actually $\Sigma(q)$, given by, in the relativistic framework \cite{Sterne2007}
\begin{equation}\label{scattering}
    |a(k,\chi)|^2=\frac{1}{k^2}\left(\Big|\sum |\kappa|e^{i\delta_\kappa(k)}\sin[\delta_\kappa(k)]P_{\ell}(\chi)\Big|^2+\Big|\sum \frac{|\kappa|}{i\kappa}e^{i\delta_\kappa(k)}\sin[\delta_\kappa(k)]P^1_\ell(\chi)\Big|^2\right),
\end{equation}
where summations are performed on the electronic states, labeled by the relativistic quantum number $\kappa$, which is related to the quantum number $\ell$ associated with the orbital momentum $L$ and with the spin $s$ by relations

\begin{equation}
    \kappa= \begin{cases}
        -(\ell+1) & \mathrm{ for }\;\;\; s=+1/2,\\
        \ell& \mathrm{ for } \;\;\; s=-1/2.
    \end{cases}
\end{equation}
The functions $P_\ell$ and $P^1_\ell$ respectively denote the Legendre and associated Legendre polynomials \cite{Abramowitz1964}. One has 
\begin{equation}
    P_{\ell}^1(x)=-(1-x^2)^{1/2}\frac{d}{dx}P_{\ell}(x).
\end{equation}
The conversion factor from atomic units to the international system of units is, for the electrical resistivity ($a_\mathrm{B}$ denotes the Bohr radius): $\frac{\hbar a_\mathrm{B}}{e^2}=21.74\,10^{-8}\,\Omega.\mathrm{m}$.

\subsection{The case of liquids}\label{sec22}

The liquid is an intermediate state of matter between solid and gas, sharing some physical properties with one or the other. Liquid densities are only slightly different from those of solids, which places them closer to the latter than to the gases. On the other hand, like gases, liquids flow, and are therefore mostly viewed as interacting gases. Until the recent developments of bright X-ray sources, it was currently admitted that liquids do not support shear-waves, \emph{i.e.}, transverse phonon modes can not exist in liquids, bringing them closer to gases. Frenkel refuted this belief in the 1950's \cite{Frenkel1955}, and asserted that shear-waves also propagate in liquids for times shorter than a mean average time characterizing the diffusion of an atom between two equilibrium positions. Frenkel's idea did not find support at his time, and it took about 50 years to be verified, thanks to the development of powerful X-ray facilities, able to investigate matter at the short time scale of the life time of transverse phonons in liquids. The existence of these transverse modes \cite{Hosokawa2009,Hosokawa2011,Hosokawa2013,Hosokawa2015,Hosokawa2021} was proved by analyzing the measured dynamic structure factors, and which support the existence of some crystalline order (formation of cages) in liquids. An X-ray diffraction experiment showed the existence of transient long-range order in melted gold \cite{Mo2018} through the coexistence of Debye-Scherrer rings and Laue diffraction peaks at times exceeding the electron-ion equilibration one.

We account for the persistence of long-range order in the liquid by adding the following correction to Ziman's resistivity \cite{Wetta2020} 
\begin{equation}\label{delta_eta}
    \delta\eta=-\dfrac{1}{3\pi {Z^*}^2 n_i}\sum_G \dfrac{N(G)}{4\pi}\mathrm{e}^{-2W(G)}
     \int_{G/2}^\infty \left(-\dfrac{\partial f}{\partial k} \right) k^2 G^2 \Bigg|a\left(k,1-\dfrac{G^2}{2k^2}\right)\Bigg|^2 \mathrm{d}k,
\end{equation}
where $N(G)$ denotes the number of reciprocal lattice vectors of the same length $G$, and $\mathrm{e}^{-2W(G)}$ the Debye-Waller factors that account for the thermal decay of the long-range order. This correction results from the extension to liquids of a prescription by Rosenfeld and Stott initially formulated for solids \cite{Rosenfeld1990}, which consists in removing the elastic (\emph{i.e.} coherent, or Bragg) scattering effects from the total structure factor used in Ziman's formula. This concept has previously been applied by Baiko {\it et al.} in the framework of astrophysics \cite{Baiko1998}. The correction vanishes as temperature grows, due to strong attenuation by the Debye-Waller factors, and also at low densities, with the function $\left(-\partial f/\partial k \right)$ then shifted to the negative energies and only contributing through its tail. Actually, $\delta\eta$ only modifies the Ziman resistivity for dense matter at moderate temperature. Indeed, in our precedent work on aluminum \cite{Wetta2020}, the correction $\delta\eta$ was essential to explain experimental liquid aluminum isobaric electrical conductivities.

\subsection{Contributions from AA models}\label{sec23}

The phase-shifts $\delta_{\kappa}(k)$ needed in Equation~(\ref{scattering}) can be obtained with the help of AA codes. The ionic structure factor $S(k)$ is obtained independently. In our model, the Ornstein-Zernicke equations are solved for charged spheres, using the Hypernetted-Chain closure relation, as described by Rogers\cite{Rogers1980}, and starting the process with the charged sphere correlation function of Held and Pignolet \cite{Held1986}. Our AA code {\sc Paradisio} was described elsewhere \cite{Wetta2018,Wetta2020}. It relies on the self-consistent calculation of the atomic structure, the one-electron states being obtained by solving the Dirac equation in the mean-field (spherical symmetry) approximation. We use the KSDT (Karasiev-Sjostrom-Dufty-Trickey) exchange-correlation (XC) functionals \cite{Karasiev2014}, with the revised parameters from Groth \emph{et al.} \cite{Groth2017}. The KSDT finite-temperature is a LDA XC functional parametrized from restricted path integral Monte Carlo data on the homogeneous electron gas (HEG) and the conventional Monte Carlo parametrization ground-state LDA XC functional (Perdew-Zunger, PZ \cite{Perdew1981}) evaluated with temperature-dependent densities. Compared to the PZ functional, the KSDT one generally lowers the direct-current (DC) electrical conductivity \cite{Karasiev2016}.

The average-atom charge $Z^*$ is not a direct output of {\sc Paradisio} (like other {\sc Inferno} based ones), and must be inferred using different charges provided by the code. This is the subject of the next section.

\section{Definition of mean ionization using AA models}\label{sec3}

\subsection{Issues in the definition of $Z^*$ within AA models}

It is often stated that ``mean ionic charge $Z^*$ is not a quantum-mechanical observable, and is therefore not clearly defined'', which suggests that this is the general case. Actually, there is a way to define a quantum-mechanical operator using the theory of scattering by a central potential \cite{Dharmawardana}. The latter must describe both the attraction of electrons neighboring an ion by the ion's nucleus and their repulsion from the charge depletion associated with the ion-ion pair distribution function. $Z^*$ is then a Lagrange multiplier that enables formal separation between bound electrons and delocalized ones, while satisfying the Friedel sum rule and ensuring charge neutrality of the system \cite{Dharmawardana2006}. Sophisticated AA models based on the neutral pseudo-atom (NPA) concept \cite{Dharma2015} meet the requirements for building this central potential, taking into account ion-ion pair correlations, and without restricting it to the ion sphere. 

Our code is based on Liberman's {\sc Inferno} model, which describes the immersion of a \emph{single ion} into a cavity (also referred to as the ion sphere, or Wigner-Seitz (WS) sphere) dug in a jellium. Moreover, the code uses the so-called A variant of this model, which only solves the Dirac equation inside the ion sphere. Ion pair correlation function is then absent from the formalism, and charge neutrality is only ensured \emph{inside} the cavity, by adjusting the chemical potential $\mu$. At no time is the model connected to Friedel's sum rule. However, as we shall see in the course of our discussions, except when pressure ionization occurs, it turns out that Friedel's rule as well as the global charge neutrality are not so badly verified even in such simple AA models. Assuming that Friedel's sum rule applies allows us to propose new definitions for $Z^*$.

\subsection{Four definitions with standard AA models}\label{sec31}

The Ziman formalism describes the scattering of a free electron gas by charged ions. Consequently, two obvious definition immediately emerge for the mean ion charge. These have been studied in detail by Murillo et al. \cite{Murillo2013}, in particular for Be and Al plasmas. First, $Z^*$ can be identified with the ideally free part of the conduction electrons, whose number reads
\begin{align}\label{Zfree}
Z_\mathrm{free}=\int_0^\infty f(\epsilon,\mu)X_\mathrm{ideal}(\epsilon) \mathrm{d}\epsilon,
\end{align}
where $X_\mathrm{ideal}(\epsilon)$ represents the ideal electron density of states (DOS):
\begin{equation*}
X_\mathrm{ideal}(\epsilon)=\dfrac{\sqrt{2\epsilon}}{\pi^2 n_i}.
\end{equation*}
$Z_\mathrm{free}$ is related to the charge density at infinity: $Z_\mathrm{free}=\overline{n}/n_i=\frac{4\pi}{3}r_\mathrm{WS}^3 \overline{n}$, where $\overline{n}$ is the jellium density and $r_\mathrm{WS}$ the ion-sphere radius. Definition $Z^*=Z_\mathrm{free}$ is fully justified as long as the real DOS behaves like the free-electron one $n(\epsilon)\varpropto \sqrt{\epsilon}$ (as for simple metals), thanks to some flexibility of Ziman's formula, which allows for compensation of small deviations from free DOS\cite{Wetta2022}. A second definition identifies $Z^*$ with the total number of continuum electrons:
\begin{align}\label{Zcont}
Z_\mathrm{cont}=\int_0^\infty f(\epsilon,\mu)X(\epsilon)\mathrm{d}\epsilon.
\end{align}
where $X(\epsilon)$ denotes the continuum DOS.

Definition (\ref{Zfree}), by considering only the charge density at infinity, excludes less extended charge densities which can nevertheless contribute to the electrical conductivity. In contrast, definition (\ref{Zcont}) suffers from the defect of including possible charges trapped in more or less broad resonances, which do not contribute to the electrical properties.

Friedel's model of an impurity (here the ion, denoted I) embedded in an ideal electron gas \cite{Friedel1952} allows us to estimate which electronic charges deviate the most from free electron-like ones. The model expresses indeed the modification of the uniform electron gas as:
\begin{equation}
Z^{\mathrm{I\,in\,UEG}}=Z^\mathrm{UEG}+Z_\mathrm{F},
\end{equation}
where the displaced charge $Z_\mathrm{F}$ is given by the Friedel sum rule, which reads, in the relativistic framework:
\begin{align}
Z_\mathrm{F}&=\dfrac{2}{\pi}\int_0^\infty \mathrm{d}\epsilon f(\epsilon,\mu)\sum_\kappa |\kappa|\dfrac{\partial\delta_\kappa (\epsilon)}{\partial\epsilon}\label{ZF1}\\ &=\dfrac{2}{\pi}\int_0^\infty \mathrm{d}\epsilon \left(-\dfrac{\partial f}{\partial\epsilon}\right)\sum_\kappa |\kappa|\delta_\kappa (\epsilon)\label{ZF2}.
\end{align}
Adding the number of displaced charges to $Z_\mathrm{free}$ yields a third possible definition for the mean ion charge:
\begin{align}\label{def3}
Z^*=Z_\mathrm{free}+Z_\mathrm{F}.
\end{align}
But, considering the equality $Z^{\mathrm{I\,in\,UEG}}=Z_\mathrm{cont}$, we have $Z^\mathrm{UEG}=Z_\mathrm{cont}-Z_\mathrm{F}$. Since electrical neutrality requires $Z^\mathrm{UEG}=Z^*$, we get the following fourth expression:
\begin{align}\label{def4}
Z^*=Z_\mathrm{cont}-Z_\mathrm{F}.
\end{align}
This last definition follows the same spirit as the one $Z^*=Z_\mathrm{cont}-Z_\mathrm{quasi-b}$ introduced by Petrov and Davidson \cite{Petrov2021} for dense plasmas when resonances appear in the continuum. $Z_\mathrm{quasi-b}$ denotes the number of electrons trapped in the resonances, considered bound, despite their positive energy. Our fourth definition generalizes Petrov and Davidson's one. Indeed, in most cases, $Z_\mathrm{F}$ remains small and only raises when some phase-shifts strongly vary with $k$, as it is the case when bound states delocalize into the continuum of energies. The corresponding partial DOS are enlarged peaks. When they are centered in the continuum, $Z_\mathrm{F}$ is positive, and the fourth definition (\ref{def4}) recovers the one of Petrov and Davidson. When the broadened bound state is still centered at negative energy, $Z_\mathrm{F}$ is negative. In a preceding paper \cite{Wetta2023}, we referred to these states as ``quasi-free'', by analogy to the ``quasi-bound'' ones which cause resonances in the continuum. In the same work, we found that definition $Z^*=Z_\mathrm{cont}-Z_\mathrm{F}$ noticeably improved agreement of our Ziman-AA conductivities with experimental ones for low density boron, aluminum, titanium and copper, for which our AA code calculated negative values for $Z_\mathrm{F}$.

\subsection{A criterion for checking Friedel's sum rule}

The continuum charge displaced by the spherical potential reads:
\begin{equation}\label{Zd}
Z_\mathrm{d}=\int_0^\infty 4\pi r^2 [n_\mathrm{cont}(r)-\overline{n}]\,\mathrm{d}r.
\end{equation}
Splitting the integral at some radius $r_c$ gives:
\begin{equation}
Z_\mathrm{d}=\int_0^{r_c} 4\pi r^2 [n_\mathrm{cont}(r)-\overline{n}]\,\mathrm{d}r + \int_{r_c}^\infty 4\pi^2 r^2 [n_\mathrm{cont}(r)-\overline{n}]\,\mathrm{d}r.
\end{equation}
For sufficiently high radius $r_c$, the second integral in the right-hand side may be neglected and:
\begin{equation}
Z_\mathrm{d}=\int_0^{r_c} 4\pi r^2 [n_\mathrm{cont}(r)-\overline{n}]\,\mathrm{d}r.
\end{equation}
In our AA model, $r_c=r_\mathrm{WS}$, and assuming that Friedel's sum rule is verified implies that $Z_\mathrm{d}=Z_\mathrm{F}$, given in Equations~(\ref{ZF1}) or (\ref{ZF2}), the phase-shifts $\delta_\kappa(k)$ being evaluated at the radius $r_\mathrm{WS}$. Checking the sum rule implies that:
\begin{equation}
Z_\mathrm{F}=\int_0^{r_\mathrm{WS}} 4\pi r^2 [n_\mathrm{cont}(r)-\overline{n}]\,\mathrm{d}r \nonumber \\
= Z_\mathrm{cont}-Z_\mathrm{free}.
\end{equation}
Our four proposals for $Z^*$ then reduce to the two following ones:
\begin{equation}
Z^*=Z_\mathrm{cont}-Z_\mathrm{F}\;(=Z_\mathrm{free})\ \text{ and }\ Z^*=Z_\mathrm{free}+Z_\mathrm{F}\;(=Z_\mathrm{cont}). 
\end{equation}
Therefore, we propose the following indicator for checking that Friedel's sum rule is reasonably verified by the AA model is:
\begin{equation}\label{critere}
\gamma=\dfrac{Z_\mathrm{cont}-Z_\mathrm{F}}{Z_\mathrm{free}+Z_\mathrm{F}}\dfrac{Z_\mathrm{cont}}{Z_\mathrm{free}}\approx 1. 
\end{equation}

In the next section, we present comparisons of our AA-based mean ionizations with the ones of two theoretical works, one (dense boron) under conditions favorable to positive $Z_\mathrm{F}$ values, and the other (low-density copper) to negative ones. In the latter case, the criterion (\ref{critere}) is also far from verified and therefore instructive.

\section{Comparison of AA mean ionizations with QMD estimates}\label{sec4}

\subsection{Case of dense boron}\label{subsec41}

\begin{figure}[!ht]
    \centering
    \includegraphics[scale=0.4,angle=0]{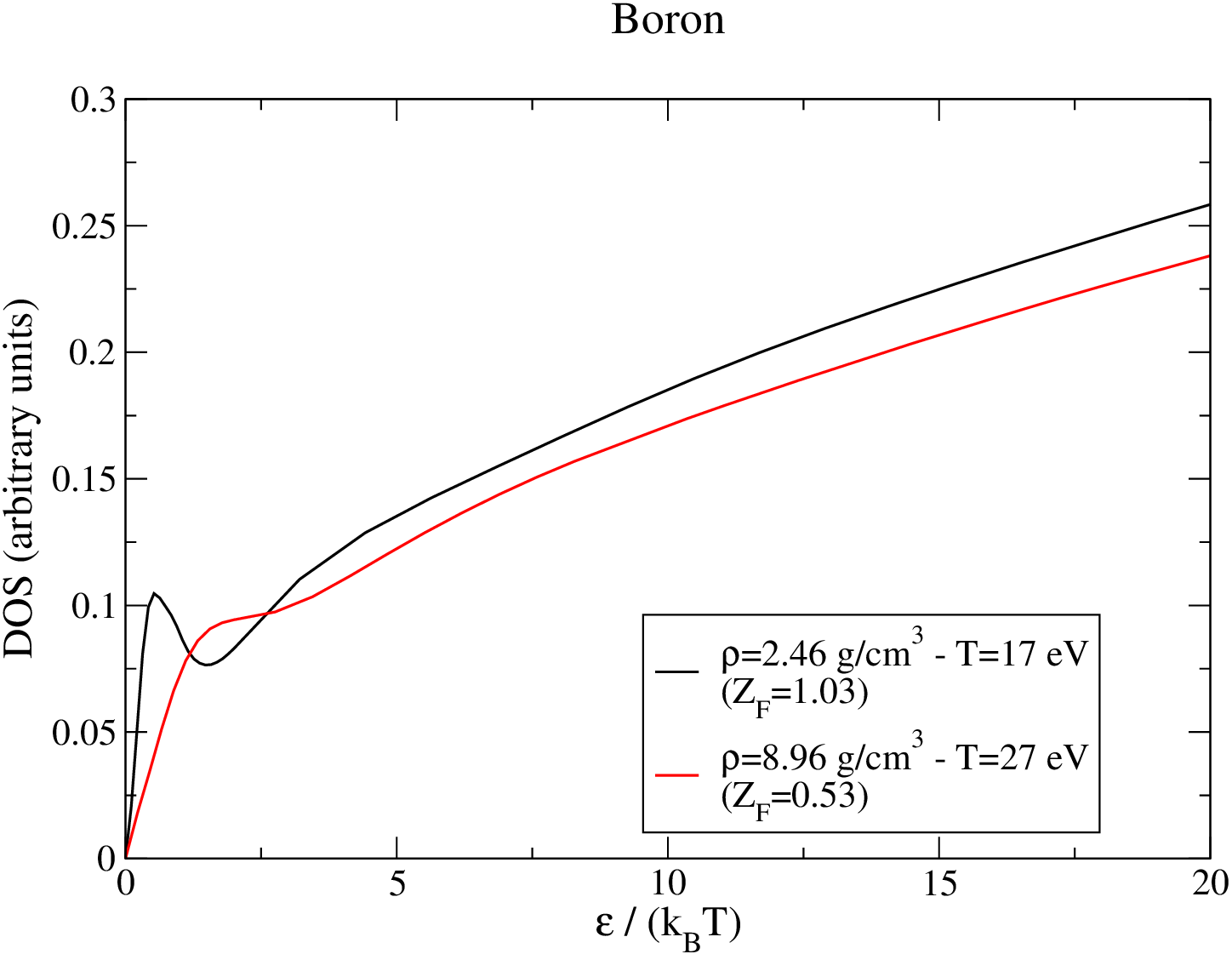}
    \caption{The figure illustrates the evolution of boron electron DOS with increasing density and temperature. Black curve: $\rho=2.46$ g.cm$^{-3}$ and $T=17$ eV, and red one: $\rho=8.96$ g.cm$^{-3}$ and $T=27$ eV. The broad resonance observed at low energies on the black curve gradually disappears as density and temperature increase, until it gives way to a quasi-UEG DOS, like that illustrated by the red curve.}
    \label{fig:Boron}
\end{figure}

In Reference~[\!\!\citenum{Blanchet2022}], Blanchet \emph{et al.} present a QMD study of boron, both at normal density and temperatures between 17 K and 697 K, and under shock compression in a comparable temperature range, including estimates of the mean ionizations from ion radial distribution functions and from QMD pressures.

The mean ionizations $Z_\mathrm{P}^*$ derived from the pressures are obtained by fitting an analytic pressure formula to the QMD values, using the following formula for the electronic contribution (in atomic units):
\begin{equation}
\frac{P_e}{n_e T} = \dfrac{1}{T} \left[T^3+3.36\,n_e\,T^{3/2}+\dfrac{9 \pi^4}{124}n_e^2 \right]^{1/3},
\end{equation}
where $n_e=Z_\mathrm{P}^* n_i$. The Nikiforov-Novikov-Uvarov formula\cite{Nikiforov2005} is an approximation of the finite-temperature Thomas-Fermi pressure, based on an expression of the Fermi integral $F_{3/2}$ in terms of powers of the integral $F_{1/2}$. The authors point out that this formula gives accurate results both for Maxwell-Boltzmann and degenerate plasmas. The ionic contribution to the pressure is taken as the perfect-gas one.\\
The authors also obtain values for the mean ion charge from their theoretical ion pair distribution functions $g(r)$. Matching the closest approach distances of OCP (one-component plasma) and the $r_{1/2}$ ones for which $g(r)=1/2$ to the QMD ones yields the effective plasma coupling parameters $\Gamma_\mathrm{eff}=\frac{Z^*_\mathrm{eff}}{r_\mathrm{ws}T}$ from which effective ion charges $Z^*_\mathrm{eff}$ can be obtained. An other way to obtain mean ionizations from pair distribution functions consists in using Ott \emph{et al.}'s parametrization \cite{Ott2014} $\Gamma_\mathrm{Ott}=-0.931+1.238 e^{1.575 r_{1/2}^3}$ for the effective plasma coupling parameter. The two latter effective values are only given for three temperatures along the solid density boron's isochore.

Under the conditions of Blanchet \emph{et al.}'s QMD studies, boron electron DOS differs from UEG one by the presence of a broad resonance superimposed on the UEG DOS ($\chi_\mathrm{ideal}(\epsilon)\varpropto\sqrt{\epsilon}$) at the lowest temperatures on the $\rho=2.46$ g.cm$^{-3}$ isochore, progressively absorbed into the UEG DOS as density and/or temperature increase (see Figure~\ref{fig:Boron}). Friedel's relation (\ref{ZF1}) then yields values of $Z_\mathrm{F}$ varying from 1.03 to 0.11 for $T=$17 eV up to 697 eV at $\rho=2.46$ g.cm$^{-3}$, and $Z_\mathrm{F}$ between 0.53 at $T=$27 eV and 0.18 at $T=$686 eV along the Hugoniot curve.

Our AA mean ionization values for boron along its normal density isochore and along boron's Hugoniot curve, obtained using the four definitions presented in the section \ref{sec3}, are respectively compared to Blanchet {et al.}'s ones in Tables~\ref{tab} and \ref{tab1}. Our best agreement with the mean ion charge $Z_\mathrm{P}$ from QMD pressures are in bold type. The last columns give an indication of the relevance of approximating the displaced charge $Z_\mathrm{d}$ by $Z_\mathrm{F}$. If this is the case, the values must be close to 1. We can see that the criterion is fairly well met. Consequently, our four definitions for $Z^*$ reduce to the two following ones: $Z^*=Z_\mathrm{cont}-Z_\mathrm{F}\approx Z_\mathrm{free}$ and $Z^*=Z_\mathrm{free}+Z_\mathrm{F}\approx Z_\mathrm{cont}$. In all cases, despite the uncertainties on the QMD extracted ionizations (Table \ref{tab} reports error bars of 10 to 20 percents on some values), we observe that $Z^*=Z_\mathrm{cont}-Z_\mathrm{F}\approx Z_\mathrm{free}$ values are the closest to the QMD ones. It is also interesting to note that the presence of an attenuated resonance on boron's DOS at 17 eV is not so detrimental as might have been feared at first sight, suggesting a greater tolerance to deviations from ideal DOS than expected.

\begin{table}[ht!]
    \centering
    \begin{tabular}{ c|c c c|c c c c|c}
          \hline
          \multicolumn{9}{c}{Boron: $\rho=2.46$ g.cm$^{-3}$} \\
          \hline
          \rule[-3ex]{0pt}{7ex} T (eV) & $Z^*_\mathrm{eff}$ & $Z^*_\mathrm{eff-Ott}$ & $Z^*_\mathrm{P}$ & $Z_\mathrm{cont}$ & $Z_\mathrm{free}$ & $Z_\mathrm{cont}-Z_\mathrm{F}$ & $Z_\mathrm{free}+Z_\mathrm{F}$ & $\gamma=\dfrac{(Z_\mathrm{cont}-Z_\mathrm{F})}{(Z_\mathrm{free}+Z_\mathrm{F})}\dfrac{Z_\mathrm{cont}}{Z_\mathrm{free}}$ \\
          \hline
          17 & $2.1 \pm 0.2$ & 1.8 & 2.04 & 3.0002 & \textbf{2.0003} & 1.9726 & 3.0279 & 0.977 \\
          86 & $3.3 \pm 0.4$ & $3.2 \pm 0.6$ & 3.43 & 3.8672 & \textbf{3.4064} & 3.6049 & 3.6687 & 1.115 \\
          174 & --- & --- & 4.30 & 4.6417 & \textbf{4.2976} & 4.4075 & 4.5318 & 1.050\\
          350 & $4.8 \pm 0.8$ & $4.8 \pm 0.8$ & 4.70 & 4.8244 & 4.7860 & \textbf{4.6433} & 4.9671 & 0.942\\
          697 & --- & --- & 4.87 & 4.9408 & 4.9248 & \textbf{4.8300} & 5.0356 & 0.962\\
          \hline
    \end{tabular}
    \caption{Boron: ionizations along the $\rho=2.46$ g.cm$^{-3}$ isochore. The table compares our mean ionization values calculated using the four definitions $Z^*=Z_\mathrm{cont},\ Z_\mathrm{free},\ Z_\mathrm{cont}-Z_\mathrm{F}$ and $Z^*=Z_\mathrm{free}+Z_\mathrm{F}$ with three ones deduced by Blanchet {\emph{et al.}} \cite{Blanchet2022} from QMD radial distribution functions ($Z^*_\mathrm{eff}$ and $Z^*_\mathrm{eff-Ott}$) and from QMD pressures ($Z^*_\mathrm{P}$). Our closest agreements with the latter appear in bold. Last column gives indication about the relevance of approximating the displaced charge $Z_\mathrm{d}$ by $Z_\mathrm{F}$ which is, in the present case, rather satisfactory.}
    \label{tab}
\end{table}

\begin{table}[ht!]
    \centering
    \begin{tabular}{c c|c|c c c c|c}
         \hline
         \multicolumn{8}{c}{Boron under shock compression}\\
         \hline
         \rule[-3ex]{0pt}{7ex} $\rho$ (g.cm$^{-3}$) & T (eV) & $Z^*_\mathrm{P}$ & $Z_\mathrm{cont}$ & $Z_\mathrm{free}$ & $Z_\mathrm{cont}-Z_\mathrm{F}$ & $Z_\mathrm{free}+Z_\mathrm{F}$ & $\gamma=\dfrac{(Z_\mathrm{cont}-Z_\mathrm{F})}{(Z_\mathrm{free}+Z_\mathrm{F})}\dfrac{Z_\mathrm{cont}}{Z_\mathrm{free}}$\\
         \hline
         8.96 & 27 & 2.48 & 3.0068 & 2.5114 & \textbf{2.4763} & 3.0419 & 0.975\\
         10.30 & 57 & 2.77 & 3.2125 & 2.8083 & \textbf{2.7950} & 3.2258 & 0.991\\
         11.57 & 144 & 3.67 & 4.1239 & 3.7225 & \textbf{3.7110} & 4.1354 & 0.994\\
         11.45 & 253 & 4.22 & 4.6155 & 4.2784 & \textbf{4.2607} & 4.6332 & 0.992\\
         11.26 & 345 & 4.47 & 4.7735 & 4.4920 & \textbf{4.4718} & 4.7937 & 0.991\\
         11.01 & 502 & 4.66 & 4.8804 & 4.7758 & \textbf{4.6446} & 5.0116 & 0.947\\
         10.83 & 686 & 4.74 & 4.9318 & 4.7657 & \textbf{4.7475} & 4.9500 & 0.993\\
         \hline
    \end{tabular}
    \caption{Boron: ionizations along the Hugoniot curve. The table compares our mean ionization values calculated using the four definitions $Z^*=Z_\mathrm{cont},\ Z_\mathrm{free},\ Z_\mathrm{cont}-Z_\mathrm{F}$ and $Z^*=Z_\mathrm{free}+Z_\mathrm{F}$ with the $Z^*_\mathrm{P}$ ones of Blanchet \emph{et al.} \cite{Blanchet2022}. Our closest agreements with the latter appear in bold. Last column gives indication about the relevance of approximating the displaced charge $Z_\mathrm{d}$ by $Z_\mathrm{F}$ which is here quite good.}
    \label{tab1}
\end{table}

\subsection{Copper plasma at density $\rho=$ 0.5 g.cm$^{-3}$}\label{subsec42}

In this case, $Z_\mathrm{QMD}^*$ was estimated by Cl\'erouin \emph{et al.} for copper at $\rho=$ 0.5 g.cm$^{-3}$ and temperatures ranging from $10^4$ K to $3 \times 10^4$ K, by fits of their QMD optical conductivities \cite{Clerouin2005}. For that purpose, the authors formulated the dielectric function as follows:
\begin{equation}
\epsilon(\omega)=1+4\pi\,n_a\,\alpha(\omega)+i \dfrac{4\pi\,\sigma(\omega)}{\omega},
\end{equation}
where $n_a$ denotes the number of atoms. Using the Lorentz oscillator model, the polarizability $\alpha(\omega)$ reads (in atomic units)
\begin{equation}
\alpha(\omega)= \sum_{j} \dfrac{1}{(\omega_j^2-\omega^2)-i\dfrac{\omega}{\tau_\mathrm{at}^j}},
\end{equation}
where the sum is running over a finite number of oscillators of frequencies $\omega_j$ and relaxation times $\tau_\mathrm{at}^j$. Finally, $\sigma(\omega)$ denotes the Drude conductivity, which is written as (in atomic units)
\begin{equation}
\sigma(\omega)= \dfrac{n_e \tau_D}{1-i\omega\tau_D}. \end{equation}
The fitting input data are the QMD DC conductivities and the frequencies of a few dominant peaks in AC ones. The relaxation times $\tau_D$ and $\tau_\mathrm{at}^j$ and the mean ionization $Z_\mathrm{QMD}^*$ are obtained using mean-square algorithms.\\ 

\begin{table}[!ht]
    \centering
    \begin{tabular}{c c|c c c c|c}
         \hline
         \multicolumn{7}{c}{Copper: $\rho=0.5$ g.cm$^{-3}$} \\
         \hline
          \rule[-3ex]{0pt}{7ex} T (K) & $Z^*_\mathrm{QMD}$ & $Z_\mathrm{cont}$  & $Z_\mathrm{free}$ & $Z_\mathrm{cont}-Z_\mathrm{F}$ & $Z_\mathrm{free}+Z_\mathrm{F}$ & 
          $\gamma=\dfrac{(Z_\mathrm{cont}-Z_\mathrm{F})}{(Z_\mathrm{free}+Z_\mathrm{F})}\dfrac{Z_\mathrm{cont}}{Z_\mathrm{free}}$\\
         \hline
          30~000 & 0.73 & 0.4880 & 0.5298 & \textbf{0.5406} & 0.4749 & 1.043\\
          25~000 & 0.56 & 0.3396 & 0.3780 & \textbf{0.4148} & 0.3029 & 1.230\\
          20~000 & 0.32 & 0.2005 & 0.2239 & \textbf{0.2779} & 0.1465 & 1.699\\
          15~000 & 0.21 & 0.0764 & 0.0865 & \textbf{0.1252} & 0.0376 & 2.940\\
          10~000 & 0.10 & 0.00793 & 0.00991 & \textbf{0.0170} & $0.867\times 10^{-3}$ & 15.7\\
          \hline
    \end{tabular}
    \caption{Comparison of our mean ionization values for copper calculated using the four definitions $Z^*=Z_\mathrm{cont},\ Z_\mathrm{free},\ Z_\mathrm{cont}-Z_\mathrm{F}$ and $Z^*=Z_\mathrm{free}+Z_\mathrm{F}$ with the $Z^*_\mathrm{QMD}$ ones of Cl\'erouin \emph{et al.} \cite{Clerouin2005}. Our closest agreement with QMD values appear in bold characters. Last column gives indication about the relevance of approximating the displaced charge $Z_\mathrm{d}$ by $Z_\mathrm{F}$ which is, in the present case, poor at low temperature, and explains the weak agreement of our results with QMD mean ionizations.} 
    \label{tab:my_label_2}
\end{table}

Table~\ref{tab:my_label_2} shows our mean ionization values for copper, calculated using the four definitions $Z^*=Z_\mathrm{cont},\ Z_\mathrm{free},\ Z_\mathrm{cont}-Z_\mathrm{F}$ and $Z^*=Z_\mathrm{free}+Z_\mathrm{F}$ with Cl\'erouin \emph{et al.} \cite{Clerouin2005} ones $Z^*_\mathrm{QMD}$. Like for boron, the last column evaluates the relevance of approximating the displaced charge $Z_\mathrm{d}$ by $Z_\mathrm{F}$. Except for $T=30\,000$ K, the values given in this column depart significantly from 1, which clearly indicates that $Z_\mathrm{d}\neq Z_\mathrm{F}$. If $Z^*=Z_\mathrm{cont}-Z_\mathrm{F}$ are still the closest to the QMD mean ionizations, they are far from matching them, even in the case of $T=30\,000$ K for which the criterion (\ref{critere}) is the best verified. However, despite our disagreement about the ionizations $Z^*$, we obtain the value $\sigma=671\ (\Omega$.cm)$^{-1}$ for the electric conductivity at $T=30\,000$ K, which is close to the QMD value 700 ($\Omega$.cm)$^{-1}$ (see Table~\ref{tab:my_label_2b}). At lower temperatures, the criterion (\ref{critere}) departs increasingly from the ideal value 1, and the conductivities deviate more and more from the QMD ones (Table~\ref{tab:my_label_2b}) However, the agreement with the QMD conductivities (see Table \ref{tab:my_label_2b}) remains satisfying, within 10 percent, for $T\geq\,20\,000$ K, when $Z^*=Z_\mathrm{cont}-Z_\mathrm{F}$ is used instead of the AA $Z_\mathrm{free}$.

\begin{table}[!ht]
    \centering
    \begin{tabular}{c c|c c c}
         \hline
         \multicolumn{5}{c}{Copper: $\rho=0.5$ g.cm$^{-3}$} \\
         \hline
          T & $\sigma_\mathrm{QMD} $ & $\sigma^{(1)}$ & $\sigma^{(2)}$ & $\sigma^{(3)}$ \\
         (K) & $(\Omega$.cm)$^{-1}$ & $(\Omega$.cm)$^{-1}$ & $(\Omega$.cm)$^{-1}$ & $(\Omega$.cm)$^{-1}$\\
         \hline
          30~000 & 700 & 655 & 546 & \textbf{671} \\
          25~000 & 540 & 383 & 436 & \textbf{487} \\
          20~000 & 310 & 205 & 232 & \textbf{297} \\
          15~000 & 200 & 67.5 & 77.7 & \textbf{118} \\
          10~000 & 100 & 7.1 & 8.9 & \textbf{15.2}\\
         \hline
    \end{tabular}
    \caption{The table compares our electrical conductivities with Cl\'erouin \emph{et al.}'s values $\sigma_\mathrm{QMD}$ \cite{Clerouin2005}. They are denoted $\sigma^{(1)},\ \sigma^{(2)}$ and $\sigma^{(3)}$, and are respectively obtained for $Z^*=Z_\mathrm{cont},\ Z_\mathrm{free}$ and $Z_\mathrm{cont}-Z_\mathrm{F}$. Our values yielding the closest agreement with QMD values are in bold type.}
    \label{tab:my_label_2b}
\end{table}

\section{Second charged-particle transport coefficient code comparison workshop: Beryllium, carbon, aluminum and gold}\label{sec5}

Tables~\ref{tab:case_Be} to \ref{tab:case_Au} report (in the second column) the conductivities obtained for beryllium, carbon, aluminum, and gold warm dense plasmas by the submitters (mentioned in the first columns) to the second charged-particle transport coefficient code comparison workshop. The last columns of the tables detail the different models being used, \emph{i.e.} AA, DFT-MD, and the KT (Kinetic Theory) one \cite{Starrett2020} which associates the quantum Landau-Fokker-Planck equation with the concept of mean-force scattering. The third columns outline the exchange-correlation functionals being applied. AA and KT methods used LDA ones, while in DFT-MD approaches GGA (generalized gradient approximation) ones are preferred.

Among the latter, the Perdew-Burke-Ernzerhof (PBE) functional owes its popularity to the fact that it is a non-empirical functional with a reasonable accuracy over a wide range of systems (empirical parametrized functionals may offer a better accuracy, but only for very specific systems). The SCAN (Strongly Constrained and Appropriately Normed) functional \cite{Sun2015} is a semi-local functional meeting all required constraints. There are indications that this functional may be better than most gradient corrected functionals \cite{Sun2016}. The rSCAN (regularized SCAN) functional \cite{Bartok2019}, introduces regularizations improving the numerics and convergence. These regularizations actually violate several of the constraints that the parent SCAN functional satisfies. However, some tests have revealed that rSCANL can be less accurate than SCAN in some cases \cite{Mejia2019}. The r$^2$SCAN (regularized-restored SCAN) functional \cite{Furness2020} modifies the rSCAN regularizations of rSCAN in order to enforce adherence to the constraints obeyed by SCAN. However, it is important to mention that it only recovers the slowly varying density-gradient expansion for exchange to second order, while SCAN recovers the expansion to fourth order. The accuracy of r$^2$SCANL is similar to the one of SCAN but with significantly improved numerical efficiency and accuracy. The SCANL \cite{Mejia2017,Mejia2018}, rSCANL, and r$^2$SCANL \cite{Mejia2020,Kaplan2022} functionals are de-orbitalized versions of SCAN, rSCAN, and r$^2$SCAN, respectively. They do not depend on the kinetic-energy density, but on the Laplacian of the density $\nabla^{2}n$ ($n$ being the electron density), instead. These functionals are available in the VASP code \cite{Kresse1996a,Kresse1996b,Kresse1999,Hafner2008}. The letter ``T'' in front of the acronym means that they apply at finite temperature. References to the models are available in the review article \cite{Stanek2024}.

The last lines of the tables present our AA results. We used the finite-temperature KSDT XC functionals \cite{Karasiev2014} with Groth \emph{et al.}'s revised parameters \cite{Groth2017} and the mean ion charge definition (\ref{def4}): $Z^*=Z_\mathrm{cont}-Z_F$. Results denoted by (A) correspond to conductivities obtained with Ziman's formula, without applying the correction $\delta\eta$ (Equation~(\ref{delta_eta})) that accounts for the persistence of solid ordering in fluid states. We found this correction negligible for beryllium and gold cases because of the high temperatures considered (respectively 4.4 eV and 10 eV). In the case of carbon at 2 eV, the correction is starting to become notable, and grows for aluminum at 1 eV. Therefore, in both latter cases, we also present results denoted by (B), including the $\delta\eta$ contribution. The type of crystal lattice considered in the formula (\ref{delta_eta}) is then specified.

Table~\ref{tab:table_Z} shows the other charges calculated with {\sc Paradisio}, for the cases discussed in this section. The case of copper will be the subject of a separate section, due to its high $Z_\mathrm{F}$ value. The last column reports the values of the ratio $\gamma$, given in Equation~(\ref{critere}).

\begin{table}[!ht]
    \centering
    \begin{tabular}{c c c c c}
    \hline
    \rule[-3ex]{0pt}{7ex} Case: & $Z_\mathrm{cont}$ & $Z_\mathrm{free}$ & $Z_\mathrm{F}$ &  
          $\gamma=\dfrac{(Z_\mathrm{cont}-Z_\mathrm{F})}{(Z_\mathrm{free}+Z_\mathrm{F})}\dfrac{Z_\mathrm{cont}}{Z_\mathrm{free}}$\\
    \hline
    Be  & 2     & 1.382  & 0.756  & 0.842 \\
    C   & 4     & 2.383  & 1.640  & 0.985 \\
    Al  & 3     & 1.847  & 1.088  & 1.058 \\
    Au  & 4.555 & 4.224  & -0.515 & 1.474 \\
    Cu  & 11    & 1.478  & 9.760  & 0.822 \\
    \hline
    \end{tabular}
    \caption{Values used for the different charges $Z_\mathrm{cont}$, $Z_\mathrm{free}$, and $Z_\mathrm{F}$, calculated with our AA code, for the cases of beryllium ($\rho=1.84$ g.cm$^{-3}$ and $T=4.4$ eV), carbon ($\rho=10$ g.cm$^{-3}$ and $T=2$ eV), aluminum ($\rho=2.7$ g.cm$^{-3}$ and $T=1$ eV), gold ($\rho=19.32$ g.cm$^{-3}$ and $T=10$ eV), and copper ($\rho=8.96$ g.cm$^{-3}$ and $T=1$ eV). This latter case is the subject of a separate section, because of its particularly high $Z_\mathrm{F}$ value. The last column reports the values of the ratio $\gamma$, defined in Equation~(\ref{critere}).}
    \label{tab:table_Z}
\end{table}

\subsection{Beryllium at $\rho=1.84$ g.cm$^{-3}$ and $T=4.4$ eV}

\begin{table}[!ht]
    \centering
    \begin{tabular}{c|c|c|c}
    \hline
        \multicolumn{4}{c}{Beryllium: $\rho=1.84$ g.cm$^{-3}$ $T=4.4$ eV}\\
    \hline
         Submitter & $\sigma$ $\left(10^4\,
         (\Omega.\mathrm{cm})^{-1}\right)$ & XC & Model\\
         \hline
G. Faussurier & 0.83 & LDA & AA\\
S.B. Hansen & 0.6$^{+0.23,-0.04}$ & LDA & AA\\
S. Hu & $0.65\pm 0.02$ & PBE & DFT-MD\\
S. Hu & $0.65\pm 0.03$ & TSCANL & DFT-MD\\
M. Sch\"orner & $0.66\pm 0.01$ & PBE & DFT-MD\\
N. Shaffer & 0.81 & LDA & KT\\
V. Sharma & $0.58\pm 0.01$ & PBE & DFT-MD\\
L. Sharma & $0.57\pm 0.01$ & PBE & (mix)-DFT-MD\\
\hline
This work (A) & 0.576 & LDA (KSDT) & AA\\
\hline
    \end{tabular}
    \caption{Case of beryllium at $T$=4.4 eV and $\rho$=1.84 g.cm$^{-3}$. Mean ion charge in our calculations: $Z^*=Z_\mathrm{cont}-Z_\mathrm{F}=1.237$.}
    \label{tab:case_Be}
\end{table}

For beryllium at $\rho=1.84$ g.cm$^{-3}$ and $T=4.4$ eV, except for Faussurier's AA and Shaffer's KT results, the theoretical values of the conductivities presented at the second charged-particle transport coefficient code comparison workshop, reported in Table~\ref{tab:case_Be}, are close to each other. DFT-MD models yield values ranging from $0.57\times 10^4\ (\Omega.\mathrm{cm})^{-1}$ to $0.66\times 10^4\ (\Omega.\mathrm{cm})^{-1}$. The AA approach of Hansen gives $0.6\times 10^4\ (\Omega.\mathrm{cm})^{-1}$, which is in excellent agreement with DFT-MD results, as our own AA work ($0.576\times 10^4\ (\Omega.\mathrm{cm})^{-1}$). The somewhat higher conductivities obtained with similar AA models result probably from different choices for the mean ion charge value $Z^*$ (not specified in the review article \cite{Stanek2024}). In our work, we used: $Z^*=Z_\mathrm{cont}-Z_\mathrm{F}=1.237$. We calculate: $\gamma=0.842$, which expresses moderate deviation from Friedel's sum rule in the present case.

\subsection{Carbon at $\rho=10$ g.cm$^{-3}$ and $T=2$ eV}

\begin{table}[!ht]
    \centering
    \begin{tabular}{c|c|c|c}
    \hline
        \multicolumn{4}{c}{Carbon: $\rho=10$ g.cm$^{-3}$ $T=2$ eV}\\
    \hline
        Submitter & $\sigma$ $\left(10^4\,
         (\Omega.\mathrm{cm})^{-1}\right)$ & XC & Model\\
         \hline
M. Bethkenhagen & $1.63\pm 0.03$ & PBE & DFT-MD\\
G. Faussurier & 1.14 & LDA & AA\\
S.B. Hansen & 1$^{+1.06,-0.36}$ & LDA & AA\\
V. Karasiev & $1.69\pm 0.06$ & PBE & DFT-MD\\
V. Karasiev & $1.04\pm 0.06$ & Tr$^2$SCANL & DFT-MD\\
C. Melton & $1.60\pm 0.01$ & LDA & DFT-MD\\
N. Shaffer & 1.87 & LDA & KT\\
V. Sharma & $1.26\pm 0.06$ & PBE & DFT-MD\\
F. Soubiran & $1.58\pm 0.04$ & PBE & DFT-MD\\
\hline
This work (A) & 0.995 & LDA (KSDT) & AA\\
This work (B-diamond) & 1.086 & LDA (KSDT) & AA\\
\hline
    \end{tabular}
    \caption{Case of carbon at $T$=2 eV and $\rho$=10 g.cm$^{-3}$. The version (B) of this work includes the subtraction of the elastic part from the structure factor assuming persistence of diamond structure. Mean ion charge in our calculations: $Z^*=Z_\mathrm{cont}-Z_\mathrm{F}=2.360$.}
    \label{tab:case_C}
\end{table}

Table~\ref{tab:case_C} shows our conductivity values for carbon at $T$=2 eV and $\rho$=10 g.cm$^{-3}$. The version (B) of this work includes the correction $\delta\eta$ (see Equation~(\ref{delta_eta})) consisting in subtracting elastic scattering by persistent crystalline sites above melting from the ion-ion total structure factor \cite{Wetta2020}. A diamond solid ordering is assumed, since solid carbon crystallizes in that structure at high compression ratio. In both (A) and (B) calculations: $Z^*=Z_\mathrm{cont}-Z_F=2.360$. 

The DFT-MD values reported in Table~\ref{tab:case_C} range from $1.04\times 10^4\ (\Omega.\mathrm{cm})^{-1}$ to $1.69\times 10^4\ (\Omega.\mathrm{cm})^{-1}$, the lowest value being obtained with the Tr$^2$SCANL XC functionals, while PBE ones yield the highest conductivities, above $1.6\times 10^4\ (\Omega.\mathrm{cm})^{-1}$. Our AA results are close to the Tr$^2$SCANL one of Karasiev, as well as Faussurier's and Hansen's AA ones, respectively, 
 $1.14\times 10^4\ (\Omega.\mathrm{cm})^{-1}$ and $10^4\ (\Omega.\mathrm{cm})^{-1}$. 
 
In the present case, we obtained: $\gamma=0.985$, which reinforces our confidence in our calculations.

\subsection{Aluminum at $\rho=2.7$ g.cm$^{-3}$ and $T=1$ eV}

\begin{table}[!ht]
    \centering
    \begin{tabular}{c|c|c|c}
    \hline
        \multicolumn{4}{c}{Aluminum: $\rho=2.7$ g.cm$^{-3}$ $T=1$ eV}\\
    \hline
         Submitter & $\sigma$ $\left(10^4\,
         (\Omega.\mathrm{cm})^{-1}\right)$ & XC & Model\\
         \hline
A. Dumi & $2.461\pm 0.004$ & LDA & DFT-MD\\
G. Faussurier & 4.18 & LDA & AA\\
S.B. Hansen & 4$^{+2.25,-1.30}$ & LDA & AA\\
V. Karasiev & $2.03\pm 0.05$ & PBE & DFT-MD\\
V. Karasiev & $1.48\pm 0.08$ & TSCANL & DFT-MD\\
G. Petrov & 5.65 & LDA & AA\\
\hline
This work (A) & 1.802 & LDA (KSDT) & AA\\
This work (B-fcc) & 2.614 & LDA (KSDT) & AA\\
This work (B-bcc) & 2.682 & LDA (KSDT) & AA\\
\hline
    \end{tabular}
    \caption{Case of aluminum at $T$=1 eV and $\rho$=2.7 g.cm$^{-3}$. Version (A) of this work uses the Ziman formalism with total ionic structure factor, and version (B) subtracts from it the effects of elastic scattering by remaining crystalline fcc or bcc ordering. Mean ion charge in our calculations: $Z^*=Z_\mathrm{cont}-Z_\mathrm{F}=1.912$.}
    \label{tab:case_Al}
\end{table}

Table~\ref{tab:case_Al} shows our conductivity values for aluminum at $T$=1 eV and $\rho$=2.7 g.cm$^{-3}$. Version (A) of this work uses the Ziman formalism using the total ion-ion structure factor, while version (B) subtracts, from the latter, the elastic scattering by a persistent crystalline (fcc or bcc) ordering. The mean ion charge obtained from our AA code is $Z^*=Z_\mathrm{cont}-Z_\mathrm{F}=1.912$. We calculate that: $\gamma=1.058$, which leads us to believe that Friedel's sum rule is fairly verified in our AA calculation.

Our results are close to DFT-MD results, especially to Dumi's calculation with LDA XC functionals, but, surprisingly, disagree by almost a factor of two with the other AA LDA ones (Faussurier, Hansen and Petrov). 

\subsection{Gold at $\rho=19.32$ g.cm$^{-3}$ and $T=10$ eV}

\begin{table}[!ht]
    \centering
    \begin{tabular}{c|c|c|c}
    \hline
    \multicolumn{4}{c}{Gold: $\rho=19.32$ g.cm$^{-3}$ $T=10$ eV}\\
    \hline
         Submitter & $\sigma$ $\left(10^4\,
         (\Omega.\mathrm{cm})^{-1}\right)$ & XC & Model\\
         \hline
G. Faussurier & 1.07 & LDA & AA\\
S.B. Hansen & 0.73$^{+0.001,-0.067}$ & LDA & AA\\
S. Hu & $1.07\pm 0.03$ & TSCANL & DFT-MD\\
V. Karasiev & $0.99\pm 0.02$ & TSCANL & DFT-MD\\
N. Shaffer & 1.18 & LDA & KT\\
F. Soubiran & $1.024\pm 0.001$ & LDA & (ext)-DFT-MD\\
\hline
This work (A) & 0.99 & LDA (KSDT) & AA\\
\hline
    \end{tabular}
    \caption{Case of gold at $T$=10 eV and $\rho$=19.32 g.cm$^{-3}$. Mean ion charge in our calculations: $Z^*=Z_\mathrm{cont}-Z_\mathrm{F}=5.070$.}
    \label{tab:case_Au}
\end{table}

Table~\ref{tab:case_Au} shows our conductivity values for gold at $T$=10 eV and $\rho$=19.32 g.cm$^{-3}$, where the mean ion charge is taken as $Z^*=Z_\mathrm{cont}-Z_\mathrm{F}=5.070$. 
In that case, the dispersion of the other results presented at the workshop is not very important. AA and DFT-MD models typically converge to $\sigma\approx 10^4 (\Omega.\text{cm})^{-1}$. Hansen's AA value and Shaffer's KT one deviate only slightly. Our own calculation agrees with the other theoretical ones reported in the table, despite the fact that we think that our evaluation of the displaced charge is poor, given the high value (close to 1.5) obtained for the ratio $\gamma$, given in Equation~(\ref{critere}). However, we observed, in sub-section \ref{subsec42}, a similar situation for copper at $\rho=0.5$ g.cm$^{-3}$ and $T\geq$ 20\,000 K, for which, however, we have calculated conductivities in good agreement with the QMD one's.

\subsection{Summary of the comparisons}

It is difficult to draw any definitive conclusion from these comparisons. As we can see, even among the same family of models (DFT-MD with PBE for instance), the dispersion of the results may be important. In some cases, our results differ from other AA LDA ones, and are closer to DFT-MD values. This is probably the consequence of, on the one hand, our efforts to bridge the gap between solid and plasma states \cite{Wetta2020} (by using an effective ionic structure factor from which elastic scattering on persistent crystalline order in liquid is removed), and, on the other, our ideas about which electrons actually contribute to conductivity \cite{Wetta2023} (leading us to subtract the ``Friedel'' charge from the continuum one).\\
The relevance of the latter definition for the mean ion charge $Z^*$ is related to the criterion given in Equation~(\ref{critere}). However, our comparisons show that the criterion is not that restrictive. Indeed, we obtained good agreement with QMD conductivities for copper at $\rho=0.5$ g.cm$^{-3}$ and $T\geq$ 20\,000 K, and with DFT-MD ones for gold at $\rho=19.32$ g.cm$^{-3}$ and $T=10$ eV, despite the fact that the criterion reaches 1.5 in these cases. This is due to some flexibility of the Ziman formula, which allows compensation of relatively small errors compared with $Z^*$, by the chemical potential values. A typical situation where such compensation fails, is the delocalization of $\ell\geq 2$ states, leading to sharp resonances in the DOS. $Z_\mathrm{F}$ is then high and comparable to $Z_\mathrm{cont}$, and even a moderate deviation of the value of the displaced charge $Z_\mathrm{d}$ from $Z_\mathrm{F}$ becomes large compared to $Z^*=Z_\mathrm{cont}-Z_\mathrm{F}$.\\
We devote the next section to the case of copper, starting with the liquid, for which resistivity has been measured at various temperatures (up to typically 2.5 times the melting one). The conclusions drawn from this study will then be applied to the calculation of the electrical conductivity at $\rho=8.96$ g.cm$^{-3}$ and $T=1$ eV.

\section{The issue of the charge neutrality in AA models: the case of copper}\label{sec6}

\subsection{Estimating the electronic charge missing in the formalism}

\begin{figure}[!ht]
    \centering
    \includegraphics[scale=0.4,angle=0]{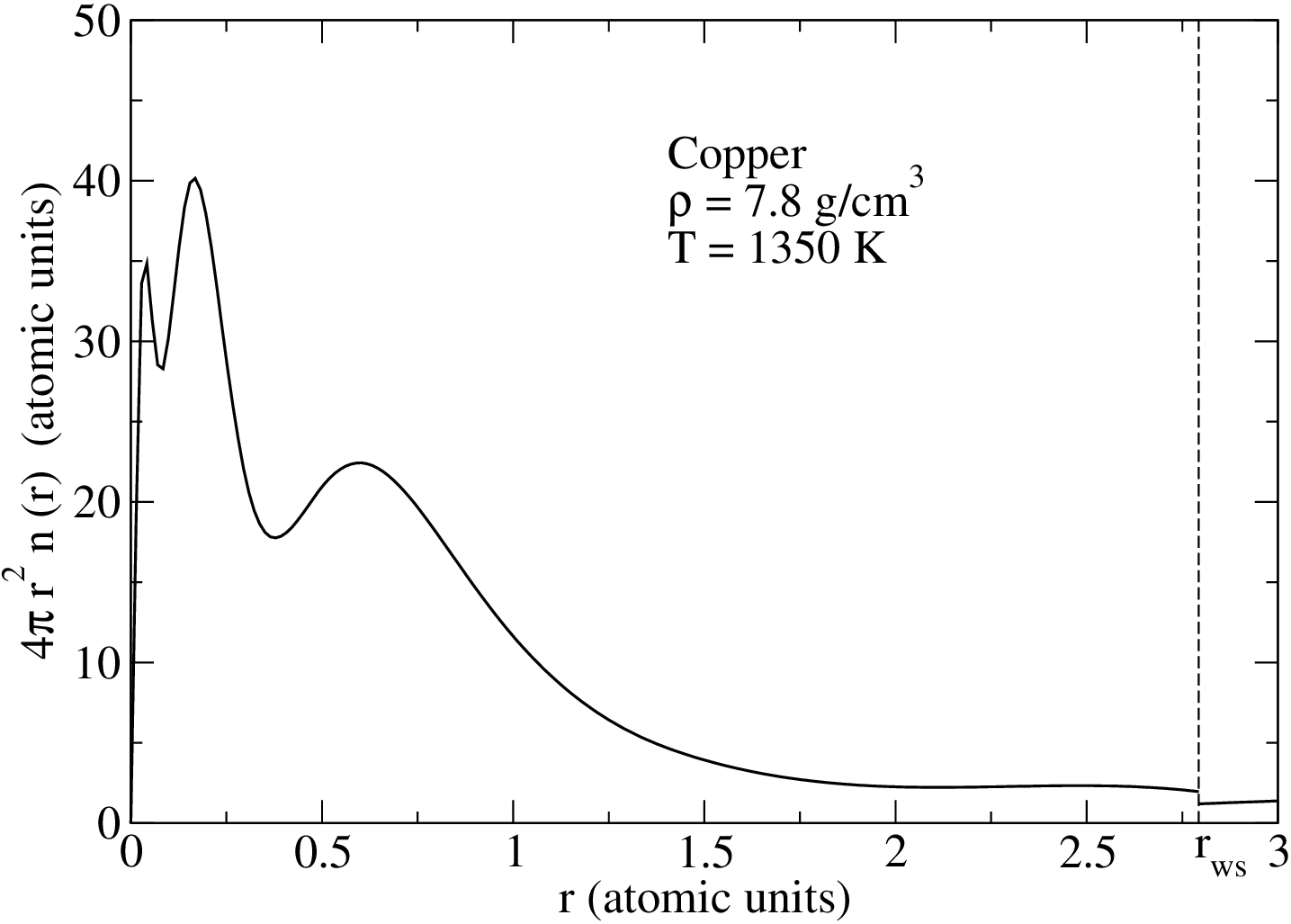}
    \caption{Charge discontinuity at the Wigner-Seitz radius. Case of liquid copper ($\rho=7.8$ g.cm$^{-3}$ and $T=$ 1350 K). }
    \label{n_r}
\end{figure}

AA models only impose charge neutrality inside the ion sphere. The continuity of the effective electron-nucleus potential is also imposed at the radius $r_\mathrm{ws}$, but it is difficult to ensure at the same time the continuity of the electron charge density at this boundary. As illustrated in Figure~\ref{n_r} for copper at melting, the charge discontinuity appears, at first glance, to be negligible. However, as we will see below, neglecting it can have a strong impact on resistivity calculation.

The total charge in the whole space is \cite{Faussurier2021}:

\begin{equation}\label{Ztot}
Z_\mathrm{tot}= (Z_\mathrm{free}+Z_\mathrm{d}) + \int_0^\infty 4 \pi r^2 n_\mathrm{bound}(r)] dr. 
\end{equation}
The first term in parentheses is the sum of the charge $Z_\mathrm{free}$ of the jellium and of the continuum charge $Z_\mathrm{d}$ displaced by the potential, given by Equation~(\ref{Zd}). The integral gives the total bound charge, including its extension outside the cavity. After separating the contributions of the ion sphere (radius $r\leq r_\mathrm{ws}$) from those of the remaining volume, and using the relation $Z=Z_\mathrm{cont}+Z_\mathrm{bound}$ that expresses the charge neutrality condition inside the ion cell, one verifies that the deviation $\delta Z=Z_\mathrm{tot}-Z$ from neutrality in the whole space is due to the fact that the electron charge density does not match the density $\overline{n}$ of the jellium at $r=r_\mathrm{ws}$:
\begin{equation}
\delta Z= \int_{r_\mathrm{ws}}^\infty 4 \pi r^2 [n_\mathrm{cont}(r)+n_\mathrm{bound}(r)-\overline{n}] dr.
\end{equation}
\emph{In order to study the interest of including this quantity in our definition of $Z^*$ for the purpose of resistivity calculations}, since our AA model cuts the electronic charge density at $r=r_\mathrm{ws}$, we must make some assumptions. We will begin with rewriting $\delta Z$ in the form:
\begin{equation}
\delta Z= \int_{r_\mathrm{ws}}^\infty 4 \pi r^2 [n_\mathrm{cont}(r)-\overline{n}] dr + \int_{r_\mathrm{ws}}^\infty 4 \pi r^2 n_\mathrm{bound}(r) dr.
\end{equation}
The first term integrates a function that oscillates around zero and progressively vanishes at infinity, and will be neglected. In fact, the same assumption is also made when the displaced charge $Z_\mathrm{d}$ is replaced by the Friedel charge $Z_\mathrm{F}$. Our previous calculations showed the relevance of this approximation for beryllium, boron, carbon, aluminum and gold plasmas. The second term integrates an exponentially decreasing function. We will assume that $(n_\mathrm{cont}(r)+n_\mathrm{bound}(r)-\overline{n})$ in the expression of $\delta Z$ can be approximated by any bell-type curve centered on $r=r_\mathrm{ws}$ and of narrow width $\delta r$ in front of the ion sphere radius. For instance, using
\begin{equation}
n_\mathrm{cont}(r)+n_\mathrm{bound}(r)-\overline{n}=\dfrac{2}{\sqrt{\pi}\,\delta r} \exp\left[-\left(\dfrac{r-r_\mathrm{ws}}{\delta r}\right)^2\right] [n_\mathrm{cont}(r_\mathrm{ws})+n_\mathrm{bound}(r_\mathrm{ws})-\overline{n}], 
\end{equation}
one obtains:
\begin{equation}
\delta Z= 4\pi r_\mathrm{ws}^2 [n_\mathrm{cont}(r_\mathrm{ws})+n_\mathrm{bound}(r_\mathrm{ws})-\overline{n}] \times \Big\{1+\dfrac{2}{\sqrt{\pi}}\left(\dfrac{\delta r}{r_\mathrm{ws}}\right) + \dfrac{1}{2} \left(\dfrac{\delta r}{r_\mathrm{ws}}\right)^2\Big\},
\end{equation}
\emph{i.e.}:
\begin{equation}
\delta Z \approx 4\pi r_\mathrm{ws}^2 [n_\mathrm{cont}(r_\mathrm{ws})+n_\mathrm{bound}(r_\mathrm{ws})-\overline{n}] +\mathrm{O}(\delta r/r_\mathrm{ws}). 
\end{equation}

Because we assumed that $\delta Z$ is localized in the vicinity of the ion sphere boundary, we will add $\delta Z$ to $Z_\mathrm{cont}$ (rather than to $Z_\mathrm{free}$, which only corresponds to charges present at infinity), yielding the following modification of the mean ion charge:
\begin{equation}
Z^*=Z_\mathrm{cont}-Z_\mathrm{F}+\delta Z,    
\end{equation}
assuming that:
\begin{equation}\label{deltaZ}
\delta Z = 4\pi r_\mathrm{ws}^2 [n_\mathrm{cont}(r_\mathrm{ws})+n_\mathrm{bound}(r_\mathrm{ws})-\overline{n}].    
\end{equation}

\begin{table}[!ht]
    \centering
    \begin{tabular}{c c c c c}
    \hline
    \rule[-3ex]{0pt}{7ex} Case: & $Z_\mathrm{cont}$ & $Z_\mathrm{F}$ & $\delta Z$ & $\dfrac{\delta Z}{Z_\mathrm{cont}-Z_\mathrm{F}}$\\
    \hline
    Be & 2     & 0.756  & 0.179 & 0.144\\
    C  & 4     & 1.640  & 0.578 & 0.245\\
    Al & 3     & 1.088  & 0.281 & 0.147\\
    Au & 4.555 & -0.515 & 0.147 & 0.029\\
    \hline
    Cu (workshop) & 11    & 9.760  & 0.742 & 0.598\\
    liquid Cu  & 11     & [10.145 , 10.306] & [0.70 , 0.75] & [0.9 , 1]\\
    \hline
    \end{tabular}
    \caption{The table reports the values obtained for $\delta Z$, using Equation~(\ref{deltaZ}), for the cases (beryllium, carbon, aluminum and gold) studied in the preceding section. The last column gives the ratio of $\delta Z$ to $Z^*=Z_\mathrm{cont}-Z_\mathrm{F}$. The case of copper, under the thermodynamic conditions adopted at the 2023 workshop, and in the liquid state for $T$ varying from melting temperature to 3500 K along the $P=0.3$ GPa isobar, are separated in order to underline the strong enhancement of the ratio in these cases.}
    \label{tab:deltaZ}
\end{table}

Table~\ref{tab:deltaZ} reports the values obtained for $\delta Z$, using Equation~(\ref{deltaZ}), for the cases (beryllium, carbon, aluminum and gold, at the densities and temperatures selected by the 2023 workshop) studied in the preceding section. The last column gives the ratio of $\delta Z$ to $Z^*=Z_\mathrm{cont}-Z_\mathrm{F}$. For these cases, the ratio remains relatively low. Bearing in mind that we are only making a rough approximation of $\delta Z$, we have decided not to take it into account in these cases (although the ratio is somewhat higher for carbon), and to reserve this option when the ratio becomes much higher. 

\begin{figure}[!ht]
    \centering
    \includegraphics[width=0.5\linewidth]{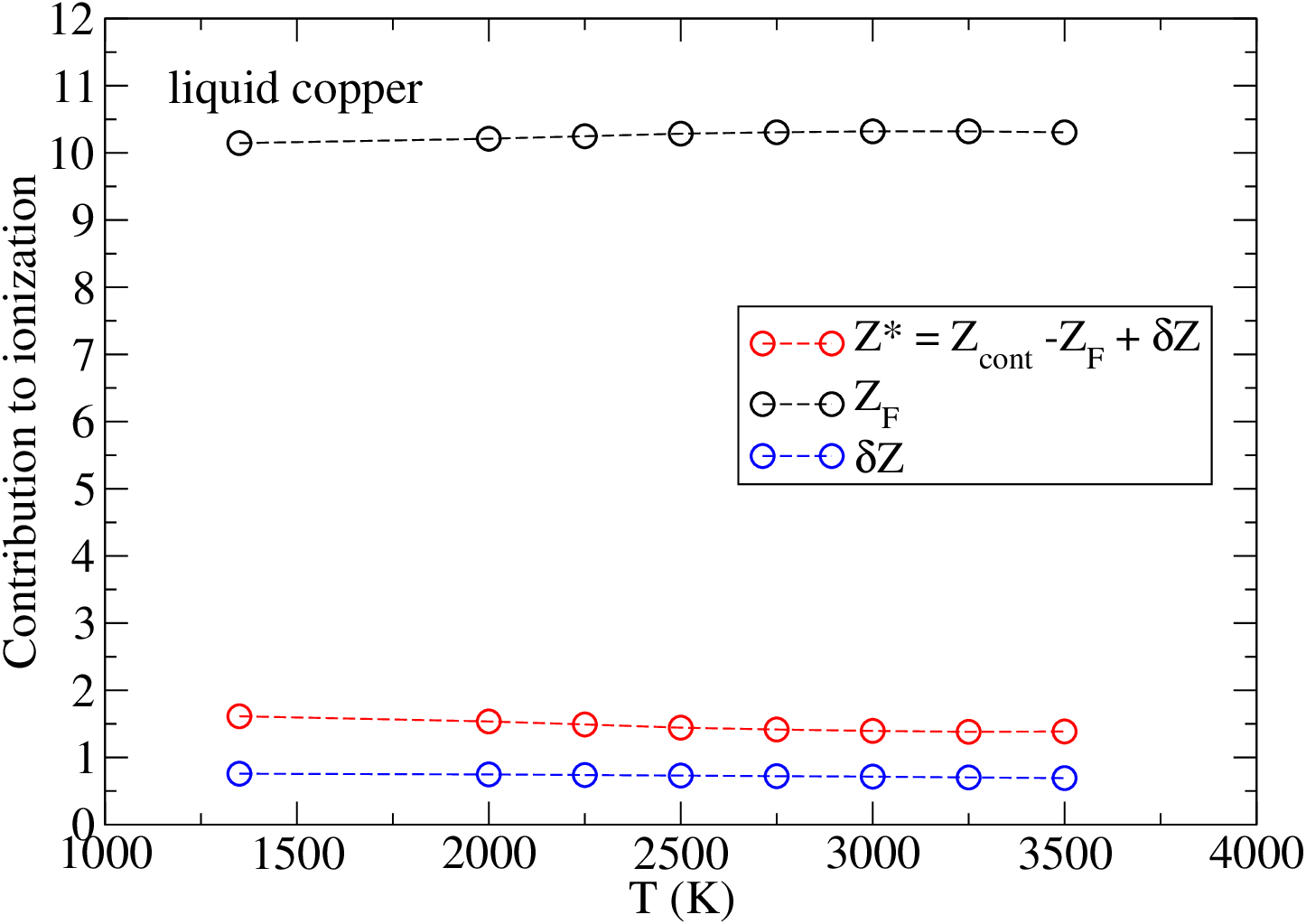}
    \caption{Liquid copper: average-atom values of $Z_\mathrm{F}$, $\delta Z$ and finally $Z^*=Z_\mathrm{cont}-Z_\mathrm{F}+\delta Z$. For all considered liquid states: $Z_\mathrm{cont}$=11. The additional charge $\delta Z$ varies from 0.76 at 1350 K (melting) to 0.69 at 3500 K, contributing to nearly the half of the value of $Z^*=Z_\mathrm{cont}-Z_\mathrm{F}+\delta Z$.}
    \label{Zstar_Cu}
\end{figure}

In the light of these values, and also of the agreement obtained between our conductivities and other theoretical results, we propose an indicator to help decide whether to include $\delta Z$ in the calculation of $Z^*$ or not. We consider that this becomes essential when:

\begin{equation}
    \dfrac{\delta Z}{Z_\mathrm{cont}-Z_\mathrm{F}}\gtrsim 0.5.
\end{equation}
This situation has arisen for copper at $\rho=8.96$ g.cm$^{-3}$ and $T=$ 1 eV, and for liquid copper, from melting temperature to 3500 K along the $P=0.3$ GPa isobar. For liquid copper, the table only shows the ranges in which $Z_\mathrm{F}$ and $\delta Z$ lie. Figure~\ref{Zstar_Cu} presents their evolution with the temperature along the isobar.

Liquid copper is a good candidate for the study of the relevance of the last redefining $Z^*=Z_\mathrm{cont}-Z_\mathrm{F}+\delta Z$ of the mean ion charge for electric conductivity calculations. Experimental resistivities are available for liquid copper from melting temperature to 3500 K \cite{Gathers1983}. In the next subsection, we will show which quantities, among $Z_\mathrm{F}$, $\delta Z$ and $\delta\eta$, are essential in the Ziman approach to explain these experiments. Finally, the conclusions that can be drawn from this study will be applied to the case of copper plasma under the thermodynamic conditions $\rho=8.96$ g.cm$^{-3}$ and $T=$ 1 eV retained at the 2023 workshop.

\subsection{Electrical resistivity of liquid copper}

Liquid copper resistivities $\eta^\mathrm{exp}$ have been measured by Gathers \cite{Gathers1983}, and are reported in column 3 of Table~\ref{tab:liq_Cu} (column 3). Columns 1 and 2 display  respectively the densities and the temperatures. Our results, using three different assumptions, are presented in columns 5, 7 and 8.

In column 5, we used $Z^*=Z_\mathrm{cont}-Z_\mathrm{F}$ (the values are given in column 4), \emph{i.e.} we neglected $\delta Z$. With this definition of the mean ion charge, and also with the other ones (among which $Z^*=Z_\mathrm{free},$ which gives values comparable to $Z^*=Z_\mathrm{cont}-Z_\mathrm{F}$), we found that the correction $\delta\eta$ is equal to zero, which is doubtful for the liquids, and our conductivities are at least four times greater than the experimental ones. Noting that none of the definitions considered so far can explain the experimental resistivities of liquid copper, we have taken into account the missing charge $\delta Z$  in the average ionic charge. In columns 7 and 8, we used $Z^*=Z_\mathrm{cont}-Z_\mathrm{F}+\delta Z$ (the values are given in column 6). Using Equation~(\ref{deltaZ}), the additional charge $\delta Z$ varies from 0.76 at 1350 K (melting) to 0.69 at 3500 K, contributing to nearly  half of the new definition $Z^*=Z_\mathrm{cont}-Z_\mathrm{F}+\delta Z$ (see Figure \ref{Zstar_Cu}). Unlike the preceding case where $\delta Z$ was neglected, the correction $\delta\eta$ is non zero, as expected for the liquids. Columns 7 and 8 present our results obtained without and with this correction. In the former case, our resistivity values are still too high, with a factor of 2.5 times the experimental ones. In the latter case, the inclusion of both $\delta Z$ and $\delta\eta$ in our modeling of the resistivity significantly improves our agreement with Gathers experiments within $\pm 15\%$.

\begin{table}[!ht]
    \centering
    \begin{tabular}{c c c | c c | c c c}
    \hline
    \multicolumn{8}{c}{Liquid copper ($Z_\mathrm{cont}=11$)}\\
    \hline
    $\rho$ & T & $\eta^\mathrm{exp}$ & $Z^*=$ & $\eta^\mathrm{AA}_\mathrm{(A)}$ & $Z^*=$ & $\eta^\mathrm{AA}_\mathrm{(A)}$ & $\eta^\mathrm{AA}_\mathrm{(B-fcc)}$
           \\
    (g.cm$^{-3}$) & (K) & ($\mu\Omega.$m) & $Z_\mathrm{cont}-Z_\mathrm{F}$ & ($\mu\Omega.$m) & $Z_\mathrm{cont}-Z_\mathrm{F}+\delta Z$ & ($\mu\Omega.$m) & ($\mu\Omega.$m)\\
    \hline
    7.832 & 1350 & 0.234 & 0.8555 & 0.954 & 1.6134 & 0.548 & 0.273\\
    7.516 & 2000 & 0.270 & 0.7892 & 1.099 & 1.5365 & 0.589 & 0.257\\
    7.319 & 2250 & 0.296 & 0.7520 & 1.225 & 1.4917 & 0.623 & 0.256\\
    7.086 & 2500 & 0.324 & 0.7140 & 1.372 & 1.4439 & 0.657 & 0.293\\
    6.921 & 2750 & 0.353 & 0.6934 & 1.508 & 1.4155 & 0.694 & 0.338\\
    6.764 & 3000 & 0.384 & 0.6807 & 1.624 & 1.3947 & 0.729 & 0.413\\
    6.565 & 3250 & 0.416 & 0.6795 & 1.795 & 1.3820 & 0.791 & 0.490\\
    6.423 & 3500 & 0.440 & 0.6942 & 1.808 & 1.3874 & 0.819 & 0.505\\
    \hline
    \end{tabular}
    \caption{Liquid copper. In column 3, $\eta^\mathrm{exp}$ are the experimental resistivities of Gathers \cite{Gathers1983}. Column 5 shows our AA resistivities calculated with $Z^*=Z_\mathrm{cont}-Z_\mathrm{F}$, given in column 4. Columns 7 and 8 report our AA values obtained with the charges $Z^*=Z_\mathrm{cont}-Z_\mathrm{F}+\delta Z$ of column 6. The subscript ``(B-fcc)'' means that correction $\delta\eta$ is applied, assuming fcc ordering, and subscript ``(A)'' that it has not. This correction is equal to zero when $\delta Z$ is neglected (columns 4 and 5), which is doubtful in the liquid state. $Z_\mathrm{cont}=11$ in all cases.}
    \label{tab:liq_Cu}
\end{table}

\begin{figure}[!ht]
    \centering
    \includegraphics[scale=0.4,angle=0]{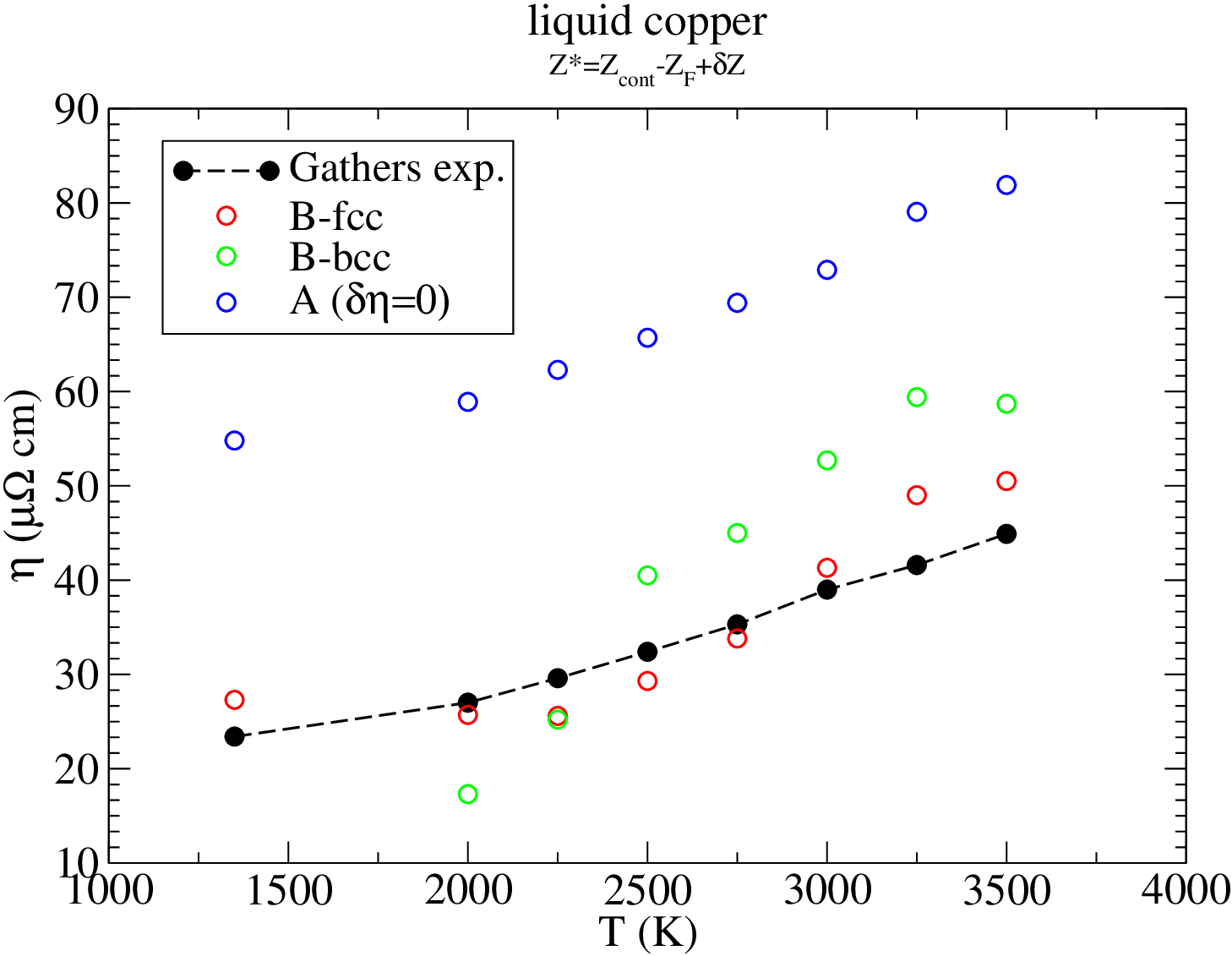}
    \caption{The figure presents the impact of different models for the correction $\delta\eta$ to the electrical resistivity. Black filled circles: experiments. Blue circles: our calculations, neglecting the correction $\delta\eta$. Green circles: B-bcc model, supposing bcc ordering in the liquid. Red circles: B-fcc model, assuming that the solid's fcc structure is transiently maintained beyond melting. In all calculations: $Z^*= Z_\mathrm{cont}-Z_\mathrm{F}+\delta Z$. The additional charge $\delta Z$ varies from 0.76 at 1350 K (melting) to 0.69 at 3500 K.}
    \label{fig:liq_Cu}
\end{figure}

For the $\delta\eta$ correction, we assumed that the remaining crystalline ordering in liquid copper is fcc (model B-fcc), \emph{i.e.} solid copper's crystal structure. We also applied the corrections obtained for bcc ordering (model B-bcc). The results are presented in Figure~\ref{fig:liq_Cu}. The black filled circles (related by a broken line) correspond to Gathers experiments. The blue circles recall the case $\delta\eta=0$ (model A, column 7 of Table \ref{tab:liq_Cu}), while the red and green ones respectively correspond to our results with B-fcc (column 8 of Table \ref{tab:liq_Cu}) and B-bcc models (not reported in the table to avoid overloading). The fact that the B-fcc model, that assumes persistence of the actual fcc solid's structure beyond melting, gives a better agreement with experiments than the B-bcc model is encouraging, and leads us to believe that our crude approximation of $\delta Z$ does not penalize our resistivity calculations so much.

This study of liquid copper is comforting enough to allow us to extend it to the case of copper at solid density and temperature $T=$ 1 eV.

\subsection{Copper at $\rho=8.96$ g.cm$^{-3}$ and $T=$ 1 eV}

\begin{table}[!ht]
    \centering
    \begin{tabular}{c|c|c|c}
    \hline
         Submitter &$\sigma$ $\left(10^4\,
         (\Omega.\mathrm{cm})^{-1}\right)$ & XC & Model\\
         \hline
K. Cochrane & 2 & PBE & DFT-MD\\
G. Faussurier & 2.75 & LDA & AA\\
S.B. Hansen & 6$^{+32.70,-2.70}$ & LDA & AA\\
S. Hu & $0.90\pm 0.02$ & PBE & DFT-MD\\
S. Hu & $1.20\pm 0.01$ & TSCANL & DFT-MD\\
G. Petrov & 3.13 & LDA & AA\\
F. Soubiran & $1.80\pm 0.02$ & LDA & DFT-MD\\
\hline
This work (A) & 2.49 & LDA (KSDT) & AA\\
This work (B-fcc) & 2.55 & LDA (KSDT) & AA\\
\hline
    \end{tabular}
    \caption{Case of copper at $T$=1 eV and $\rho$=8.96 g.cm$^{-3}$. The version (B-fcc) of this work includes the correction $\delta\eta$ that consists to subtract the elastic scattering by transient fcc ordering from the structure factor, while version (A) neglects it. Mean ion charge in our calculations: $Z^*=Z_\mathrm{cont}-Z_\mathrm{F}+\delta Z=1.983.$}
    \label{tab:case_Cu}
\end{table}

The theoretical electrical conductivities submitted at the 2023 workshop are reported in the upper part of the Table~\ref{tab:case_Cu}. They vary from $0.9\times 10^4$ to $6\times 10^4\,(\Omega.\text{cm})^{-1}$. The DFT-MD studies yield the lowest values ($\sigma\lesssim 2\times 10^4\,(\Omega.\text{cm})^{-1}$), whereas the AA models give electrical conductivities over $2.75\times 10^4\,(\Omega.\text{cm})^{-1}$. To be consistent with our previous study on liquid copper, which highlighted the need to include the additional charge $\delta Z$ in the mean ion charge $Z^*$, and to apply the correction $\delta\eta$ to the Ziman resistivity, in order to explain the liquid's resistivities, we use both in our calculations. Our results are given in the last two rows of Table~\ref{tab:case_Cu}, and are close to the AA calculations of Faussurier. Model (A) neglects the correction $\delta\eta$. The difference from the model (B-fcc), which takes it into account, is, as expected, quite negligible at the temperature $T=1$ eV. 

For information, we obtained $\sigma=1.59\times 10^4\,(\Omega.\text{cm})^{-1}$ with $Z^*=Z_\mathrm{cont}-Z_\mathrm{F}=1.241$, which also lies in the range of values calculated with DFT-MD and other AA approaches. This suggests that, like the contribution $\delta\eta$, the impact of the additional charge $\delta Z$ decreases with increasing temperature. However, the option $\delta Z=0$, although it also agrees with other theoretical results, breaks the continuity between our description of the liquid state and the copper plasma.

\section{Conclusions}

In this work, we presented calculations of electrical resistivities for beryllium, boron, carbon, aluminum, copper and gold, within the Ziman-Evans theory, using  our average-atom code {\sc Paradisio}, for thermodynamic conditions likely to yield significant differences in the mean ionic charge $Z^*$ according to its meaning. We compared and discussed different ways of characterizing the mean ion charge (or mean ionization), in order to figure out which electrons really contribute to electric conduction. Our results were compared to experimental data, and to published theoretical values, in particular from the second charged-particle transport coefficient code comparison workshop, which was held in July 2023 at Lawrence Livermore National Laboratory.

For dense plasmas, the number $Z_\mathrm{F}$ (referred to as the ``Friedel charge'' above) of electrons that are the most strongly scattered by the central potential, rises significantly in the conditions where resonances occur in the DOS. $Z_\mathrm{F}$ is then large and positive, and definition $Z^*=Z_\mathrm{cont}-Z_\mathrm{F}$ subtracts the number of electrons trapped in the resonance, considered as ``quasi-bound'', from the total number $Z_\mathrm{cont}$ of continuum electrons. The definition $Z^*=Z_\mathrm{cont}-Z_\mathrm{F}$ is actually flexible and applies successfully to different plasma conditions: indeed, $Z_\mathrm{F}$ can be $>0$ or $<0$, depending whether the potential is attractive or repulsive. The latter case occurs for low-density plasmas. Some bound electrons can be considered as  ``quasi-free'' and are included in the number of electrons contributing to conductivity. This definition of the mean ionization (or more exactly of the electrons effectively contributing to the electrical conduction), gives in most cases the best agreement with measurements and quantum-molecular-dynamics simulations. 

Definition $Z^*=Z_\mathrm{cont}-Z_\mathrm{F}$ lies on the assumption that deviations from the Friedel sum rule are limited, which is a prerequisite for the charge displaced by the potential to be approximated by $Z_\mathrm{F}$. We have proposed an indicator (the ratio $\gamma$ given in Equation~(\ref{critere})) to check the validity of this approximation, which can be easily calculated using the electronic charges provided by the code. However, because of the ability of the Ziman formula to compensate a reasonable inaccuracy of $Z^*$ by the chemical potential, deviations of $\gamma$ from the ideal value $\gamma=1$ (which indicates the exact verification of the sum rule) must be important for impacting the conductivity calculation. Concretely, we only observed strong disagreement of our AA calculations with QMD ones for expanded copper at low temperature, in which case $\gamma\gtrsim$ 2.

Finally, we found that a ``missing charge'' $\delta Z$ due to deviation from neutrality in the whole space must, in some cases, be taken into account in the most reliable way. Departure from neutrality is related to the fact that, in our average-atom model, the atom is confined in the Wigner-Seitz sphere, while its environment is replaced by a jellium. The impact of the deviation, and consequently the value of the ``missing charge'', are linked to the charge discontinuity jump at the sphere's boundary that results from this assumption.  In order to estimate the impact of $\delta Z$ on $Z^*$, we proposed an indicator, using the charge discontinuity. Ideally, its value must be as low as possible to allow us to neglect $\delta Z$ in the definition of $Z^*$. High values are possible for ``d-block'' metals, particularly in their liquid state. Indeed, we showed, in the case of liquid copper, that the inclusion of $\delta Z$ in the definition of the mean ion charge is necessary to achieve agreement of our AA resistivities with experimental values. The importance of accounting for a persisting crystalline order in the liquid, which is a specificity of our approach, appears also clearly in the light of our study of liquid copper.

Finally, we would like to point out that our definitions including the notions of displaced and/or missing charges tend towards $Z_\mathrm{free}$  when the real DOS is not too far from that of the free electron gas, which is notably the case for simple metals (Al, Be\dots). We demonstrated that this still holds true in the presence of an attenuated resonance superimposed on a free-electron-like DOS, similar to that observed for boron at 17 eV, which extends the range of validity of the AA $Z_\mathrm{free}$ for conductivity calculations. Our new definition of the mean-ion charge does not contradict the intuitive approximation of $Z^*$ by $Z_\mathrm{free}$, which remains valid in many situations \cite{Murillo2013}. It merely aims at accounting for charges that, in some cases, are not reflected in the value derived from the electron charge density at infinity.

\clearpage

\appendix

\section*{Acknowledgments}

We are indebted to Chandre Dharma wardana for fruitful discussions about various aspects of the calculation of the DC conductivity within Ziman's formulation. 


\providecommand{\url}[1]{\texttt{#1}}
\providecommand{\urlprefix}{}
\providecommand{\foreignlanguage}[2]{#2}
\providecommand{\Capitalize}[1]{\uppercase{#1}}
\providecommand{\capitalize}[1]{\expandafter\Capitalize#1}
\providecommand{\bibliographycite}[1]{\cite{#1}}
\providecommand{\bbland}{and}
\providecommand{\bblchap}{chap.}
\providecommand{\bblchapter}{chapter}
\providecommand{\bbletal}{et~al.}
\providecommand{\bbleditors}{editors}
\providecommand{\bbleds}{eds: }
\providecommand{\bbleditor}{editor}
\providecommand{\bbled}{ed.}
\providecommand{\bbledition}{edition}
\providecommand{\bbledn}{ed.}
\providecommand{\bbleidp}{page}
\providecommand{\bbleidpp}{pages}
\providecommand{\bblerratum}{erratum}
\providecommand{\bblin}{in}
\providecommand{\bblmthesis}{Master's thesis}
\providecommand{\bblno}{no.}
\providecommand{\bblnumber}{number}
\providecommand{\bblof}{of}
\providecommand{\bblpage}{page}
\providecommand{\bblpages}{pages}
\providecommand{\bblp}{p}
\providecommand{\bblphdthesis}{Ph.D. thesis}
\providecommand{\bblpp}{pp}
\providecommand{\bbltechrep}{}
\providecommand{\bbltechreport}{Technical Report}
\providecommand{\bblvolume}{volume}
\providecommand{\bblvol}{Vol.}
\providecommand{\bbljan}{January}
\providecommand{\bblfeb}{February}
\providecommand{\bblmar}{March}
\providecommand{\bblapr}{April}
\providecommand{\bblmay}{May}
\providecommand{\bbljun}{June}
\providecommand{\bbljul}{July}
\providecommand{\bblaug}{August}
\providecommand{\bblsep}{September}
\providecommand{\bbloct}{October}
\providecommand{\bblnov}{November}
\providecommand{\bbldec}{December}
\providecommand{\bblfirst}{First}
\providecommand{\bblfirsto}{1st}
\providecommand{\bblsecond}{Second}
\providecommand{\bblsecondo}{2nd}
\providecommand{\bblthird}{Third}
\providecommand{\bblthirdo}{3rd}
\providecommand{\bblfourth}{Fourth}
\providecommand{\bblfourtho}{4th}
\providecommand{\bblfifth}{Fifth}
\providecommand{\bblfiftho}{5th}
\providecommand{\bblst}{st}
\providecommand{\bblnd}{nd}
\providecommand{\bblrd}{rd}
\providecommand{\bblth}{th}

\end{document}